\documentclass[sigconf]{acmart}

\usepackage{booktabs} 

\copyrightyear{2022}
\acmYear{2022}
\setcopyright{acmcopyright}\acmConference[WWW '22]{Proceedings of the ACM Web
Conference 2022}{April 25--29, 2022}{Virtual Event, Lyon, France}
\acmBooktitle{Proceedings of the ACM Web Conference 2022 (WWW '22), April
25--29, 2022, Virtual Event, Lyon, France}
\acmPrice{15.00}
\acmDOI{10.1145/3485447.3511939}
\acmISBN{978-1-4503-9096-5/22/04}

\usepackage{amsthm}
\usepackage{amscd,amsfonts,amsbsy,rotating}
\usepackage{balance}
\usepackage{graphicx}
\usepackage{epsfig,epstopdf}
\usepackage{subfigure}
\usepackage{multirow}
\usepackage{booktabs}
\usepackage{color,xcolor}
\usepackage{url}
\usepackage{latexsym,bm}
\usepackage{enumitem,balance,mathtools}
\usepackage{wrapfig}
\usepackage{euscript}
\usepackage{ifpdf}
\usepackage{diagbox}
\usepackage{caption}
\usepackage{makecell}
\usepackage{subfigure}
\usepackage[leftcaption]{sidecap}
\usepackage{textcomp}
\usepackage[normalem]{ulem}
\usepackage{breakurl}

\usepackage{array}
\newcolumntype{N}{@{}m{0pt}@{}}

\usepackage{lipsum}
\PassOptionsToPackage{hyphens}{url}\usepackage{hyperref}

\usepackage[ruled,linesnumbered]{algorithm2e}

\usepackage{algorithmic}

\newtheorem{definition}{Definition}

\newcommand{\minisection}[1]{\vspace{5pt}\noindent\textbf{#1.}}

\begin{document}
\title{Who to Watch Next: Two-side Interactive Networks \\ for Live Broadcast Recommendation}
	\author{
	Jiarui Jin$^{1,*}$, Xianyu Chen$^{1,*}$, Yuanbo Chen$^{2,*}$, Weinan Zhang$^{1,\dag}$, Renting Rui$^{1}$,\\ Zaifan Jiang$^{2}$, Zhewen Su$^{2}$, Yong Yu$^{1}$.}
\affiliation{$^1$Shanghai Jiao Tong University, China; $^2$Alibaba Group, China}
\email{{jinjiarui97, xianyujun, wnzhang, ruirenting, yyu}@sjtu.edu.cn, {yuanbo.cyb,zaifan.jzf, zhewen.su}@alibaba-inc.com}

    \renewcommand{\shortauthors}{J. Jin, et al.}
	\renewcommand{\shorttitle}{TWINS}
	\settopmatter{printacmref=false}


\begin{abstract}
	With the prevalence of live broadcast business nowadays, a new type of recommendation service, called live broadcast recommendation, is widely used in many mobile e-commerce Apps.
	Different from classical item recommendation, live broadcast recommendation is to automatically recommend user anchors instead of items considering the interactions among triple-objects (i.e., users, anchors, items) rather than binary interactions between users and items.
	Existing methods based on binary objects, ranging from early matrix factorization to recently emerged deep learning, obtain objects' embeddings by mapping from pre-existing features.
	Directly applying these techniques would lead to limited performance, as they are failing to encode collaborative signals among triple-objects.
	In this paper, we propose a novel 
	\textbf{TW}o-side \textbf{I}nteractive \textbf{N}etwork\textbf{S} (\textbf{TWINS}) for live broadcast recommendation.
	In order to fully use both static and dynamic information on user and anchor sides, we combine a product-based neural network with a recurrent neural network to learn the embedding of each object. 
	In addition, instead of directly measuring the similarity, TWINS effectively injects the collaborative effects into the embedding process in an explicit manner by modeling interactive patterns between the user's browsing history and the anchor's broadcast history in both item and anchor aspects.
	Furthermore, we design a novel co-retrieval technique to select key 
	items among massive historic records efficiently. 
	Offline experiments on real large-scale data show the superior performance of the proposed TWINS, compared to representative methods;
    and further results of online experiments on Diantao App show that TWINS gains average performance improvement of around 8\% on ACTR metric, 3\% on UCTR metric, 3.5\% on UCVR metric.
	\end{abstract}
	\settopmatter{printacmref=false} 

\maketitle

{
\renewcommand{\thefootnote}{\fnsymbol{footnote}}
\footnotetext[1]{Jiarui Jin, Xianyu Chen, and Yuanbo Chen contributed equally to the work.}
\footnotetext[2]{Weinan Zhang is the corresponding author.}
}

{\fontsize{8pt}{8pt} \selectfont
	\textbf{ACM Reference Format:}\\
	Jiarui Jin, Xianyu Chen, Yuanbo Chen, Weinan Zhang, Renting Rui, Zaifan Jiang, Zhewen Su, Yong Yu. 2022. Who to Watch Next: Two-side Interactive Networks for Live Broadcast Recommendation. In \textit{Proceedings of the ACM Web Conference 2022 (WWW '22), April 25--29, 2022, Lyon, France} ACM, New York, NY, USA, 10 pages.
	\url{https://doi.org/10.1145/3485447.3511939}}
\vspace{-1mm}

\section{Introduction} 
\begin{figure}[t]
	\centering
	\includegraphics[width=1.0\linewidth]{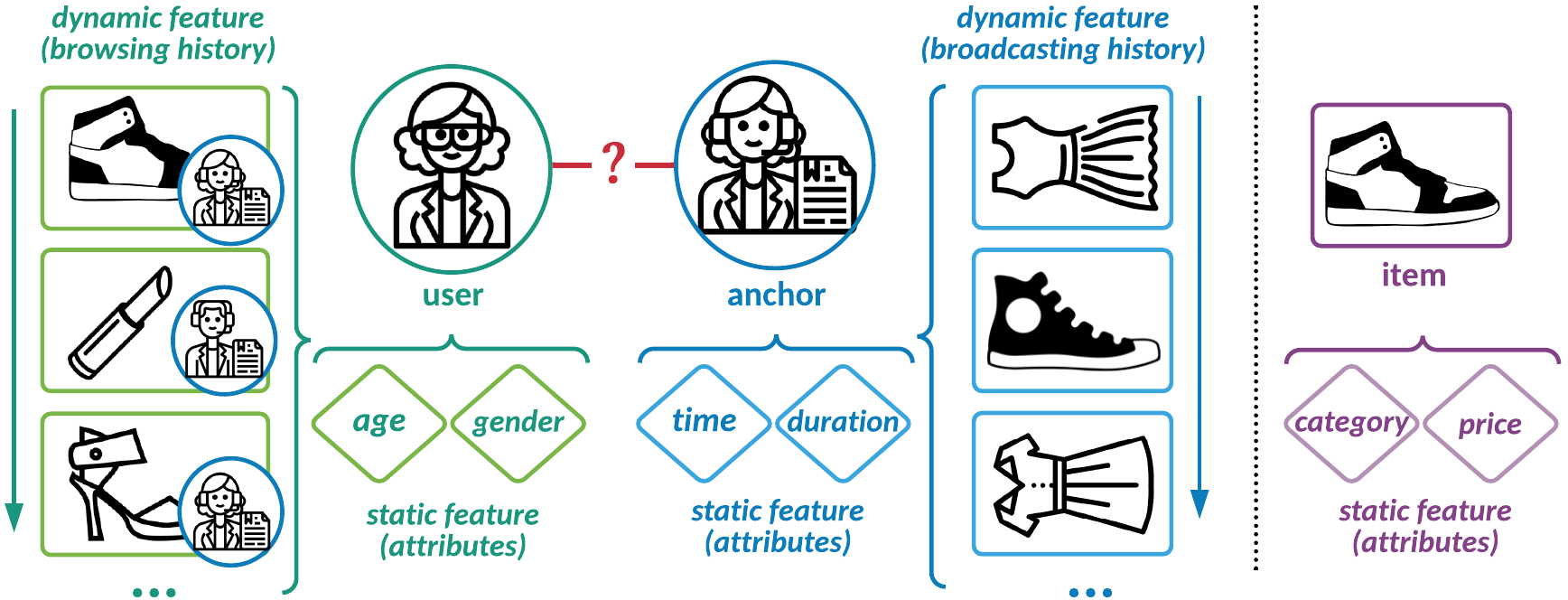}
	\vspace{-6mm}
	\caption{
		An illustrated live broadcast recommendation example for recommending an anchor to a user (while there will be a list of anchors in real-world Apps).
	}
	\label{fig:task}
	\vspace{-6mm}
\end{figure}

With the establishment of mobile Internet, the focus of e-commerce has moved from personal computers to smartphones, which significantly encourages the emergence and development of live broadcast services.
Live broadcast recommendation has become popular, especially in the past two years, because of anchors' revealing selection and expressiveness powers, as such, free users from tedious searching and comparing in mobile phones.
Figure~\ref{fig:task} shows a live broadcast recommendation example.
According to historical information on user and anchor sides, a list of appropriate anchors will be automatically generated for a user.
In a live broadcast recommendation system, the historical information can be roughly categorized into two types for both two-fold.
The first one is static data, containing attribute information such as user and anchor profiles.
The other one is dynamic data, containing user browsing history represented as triple interactions (i.e., $\langle \text{users, anchors, items} \rangle$) and broadcasting history represented as binary interactions (i.e., $\langle \text{anchors, items} \rangle$).

Notably, the live broadcasting recommendation here is significantly different from existing recommendation tasks in the following aspects:
(1) Different from traditional recommendations of query \citep{zhang2006mining,cao2008context} or item \citep{jin2020efficient,zhou2018deep} and recently introduced intent recommendation \citep{fan2019metapath}, it recommends anchors instead of queries or items to users.
(2) Our live broadcast recommendation needs to consider the interactions among triple-objects (i.e., users, anchors, and items) rather than binary interactions between users and items.
(3) Different from queries and items, the status of anchors (i.e., broadcasting or not) always changes frequently.

Existing methods for live broadcast recommendation employed in industry, such as Taobao and Kuaishou, usually extract handcrafted features in user and anchor sides, and then feed these features to a classifier ranging from early matrix factorization \citep{koren2009matrix} to recently emerged deep learning \citep{cheng2016wide}.
These approaches heavily rely on laboring feature engineering and fail to use the rich, dynamic interactions among objects fully.
However, as the anchors are rapidly changing the items sold in the broadcast room, it is really critical to model their interactions to capture the temporal behaviors.
Moreover, their techniques proposed for binary-objects (i.e., users, items), obtaining an object's embedding by mapping from pre-existing features; are indirectly to extend to encode collaborate signals among triple-objects (i.e., users, anchors, items).

In summary, we introduce a recently emerged, but seldom exploited, live broadcast recommendation problem; to handle which, we are at least required to address the following challenges:
\begin{itemize}[topsep = 3pt,leftmargin =5pt]
\item \textbf{(C1)} How to build correlations between users and anchors, since their relevance is an inherent attribute hidden in complex static and dynamic features in both user and anchor sides?
Consider the following scenario (shown in Figure~\ref{fig:task}).
When the teen opens a live broadcast recommendation App, the recommender system returns several anchors based on her profile and historical data.
For each pair of user and anchor, we are required to model two kinds of features; namely static ones often formed as categorical attributes (e.g., user's gender and anchor's broadcast time), and dynamic ones often formulated as sequential data (e.g., user's browsing history and anchor's broadcasting history). 
These historical data consist of related items which also have their own static features (e.g., item's price and brand).
\item \textbf{(C2)} How to capture the collaborative signals between user and anchor sides?
As stated in \citep{wang2019neural}, the mainstream methods for recommendation, either early employed shallow or recently proposed deep models, fail to capture the hidden collaborative information.
Further analysis in \citep{qu2019end,jin2020efficient} reveals the early summarization issue exists in the structural data, and we argue that similar issue occurs here, where existing approaches usually compress all the information together in each side regardless of rich interactive (i.e., `AND') patterns between user's browsing and anchor's histories.
Take Figure~\ref{fig:task} as an instance.
The motivation of the teen entering the live broadcast room can come from the current anchor selling the item she watched before, which can be modeled by \textbf{AND} operation over anchor broadcast items and user browsed items.
\item \textbf{(C3)} How to distinguish the key information and filter out the noise?
Recent works \citep{ren2019lifelong,pi2020search} 
reveal that
observe that long-term dependencies exist in the historical records.
However, since the length of historical 
sequences vary for different users due to diverse activeness or registration time and some of them are extreme long, it is not practical to maintain the whole behavior history of each user for real-time online inference.
\end{itemize}

In this paper, we propose a novel \textbf{TW}o-side \textbf{I}nteractive \textbf{N}etwork\textbf{S} (\textbf{TWINS}) for live broadcast recommendation.
In seeking for a proper way to effectively capture correlations between user and anchor according to complicated behavior histories in these two sides, we introduce a new two-side network architecture, where we combine product-based neural network (PNN) \citep{qu2016product} and recurrent neural network (RNN) \citep{hochreiter1997long} in each side to simultaneously model static and dynamic features.
Concretely, for static features usually formed as categorical data, we establish an embedding vector for each category and adopt PNN to capture the hidden interactive patterns; and then incorporate it with contextual information by feeding the learned embeddings into the RNN model (\textbf{C1}).
A principal way to discover the hidden collaborative signal is to employ collaborative filtering methods such as SVD++ \citep{koren2008factorization};
however, these techniques still suffer from the early summarization issue and cannot be directly applied to live broadcast recommendation scenarios.
Hence, we first propose interaction networks to measure the similarity of user and anchor in the two-side architecture, in both item and anchor aspects, which are further aggregated to form our final objective (\textbf{C2}). 
To efficiently handle long-sequential data,
inspired by the recently proposed retrieval model \citep{pi2020search,qin2020user}, we design a novel co-retrieval mechanism to search and retrieve the relevant items in user and anchor sides (\textbf{C3}).

We conduct thorough experiments on four real-world datasets to verify the superiority of TWINS over recent state-of-the-art methods. 
Further, TWINS has been deployed on the recommender system of a mainstream Diantao App, where the online A/B test shows that TWINS achieves better performance than baseline methods on all the measurement metrics.

\begin{figure*}[t]
	\centering
	\includegraphics[width=1.0\linewidth]{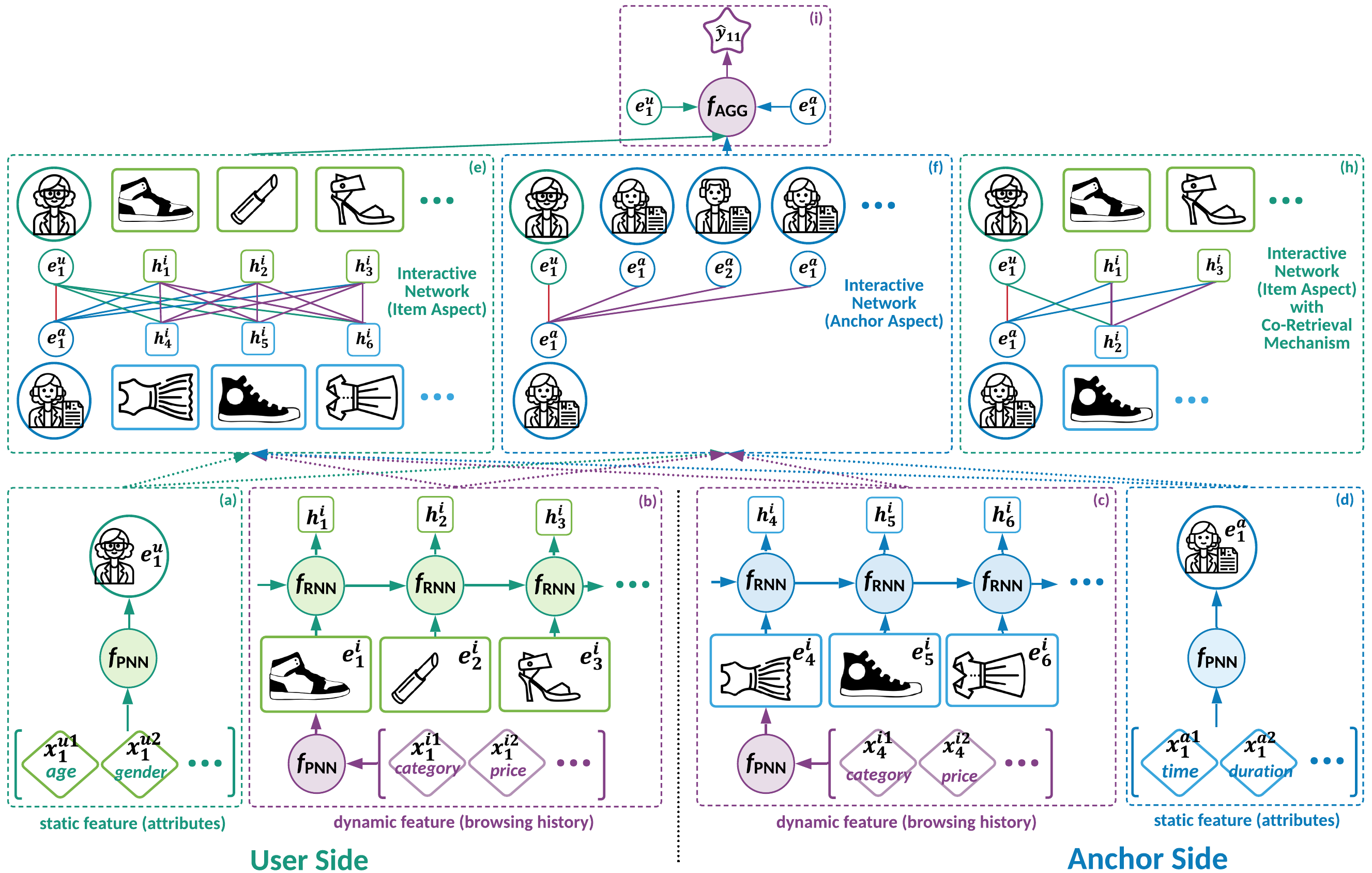}
	\vspace{-6mm}
	\caption{
		The overview of TWINS.
		The bottom part (i.e., (a)-(d)) shows the two-side architecture where we use the PNN to encode the static (categorical) attributes and the RNN to encode the dynamic (sequential) histories 
		in user and anchor sides.
		The up part (i.e., (e)-(g)) illustrates the interaction networks which take these embedding vectors as input and outputs the interactive patterns from item and anchor aspects, which are further aggregated with user and anchor static features to make the final prediction (i.e., $\widehat{y}_{11}$ for user $u_1$ and anchor $a_1$).
		Moreover, we design a co-retrieval mechanism, as illustrated in (h), to decrease the interaction computations by selecting a subset of historical items for the interaction instead of the whole set. 
	}
	\label{fig:overview}
	\vspace{-3mm}
\end{figure*}

\section{Preliminaries}
\subsection{Problem Formulation}
\label{sec:problem}
We begin by describing the live broadcast recommendation task and associated notations.
\begin{definition}
\textbf{Live Broadcast Recommendation.}
Given a triplet $\langle \mathcal{U}, \mathcal{A}, \mathcal{I} \rangle$, where $\mathcal{U}=\{u_1,\ldots,u_P\}$ denotes the set of $P$ users, $\mathcal{A}=\{a_1,\ldots,a_N\}$ denotes the set of $N$ anchors, and $\mathcal{I}=\{i_1,\ldots,i_Q\}$ denotes the set of $Q$ items. 
The purpose of live broadcast recommendation is to recommend the most related anchor $a\in \mathcal{A}$ to a  user $u\in\mathcal{U}$ according to the user's browsing history $\mathcal{H}^u$ and the anchor's broadcasting history $\mathcal{H}^a$.
\end{definition}
In our application, as shown in Figure~\ref{fig:task}, a user's (e.g., the $p$-th user's) browsing history $\mathcal{H}^u_p$ is constituted by a series of her visited items with associated anchors.
For convenience, we establish a set of user's browsed items denoted as $\mathcal{H}_p^{ui}$ and the other set of user's visited anchors denoted as $\mathcal{H}_p^{ua}$.
We then have $\mathcal{H}_p^u=\mathcal{H}_p^{ui}\cup \mathcal{H}_p^{ua}$ where $\mathcal{H}_p^{ui} \subseteq \mathcal{I}$, $\mathcal{H}_p^{ua} \subseteq \mathcal{A}$; and similarly an anchor's (e.g., the $n$-th anchor's) broadcasting history $\mathcal{H}^a_n=\mathcal{H}_n^{ai} \subseteq \mathcal{I}$ holds for any $p\in\{1,\ldots,P\}$ and any $n\in\{1,\ldots,N\}$.
Besides the complex interactions among these triplet-objects (a.k.a., dynamic feature in Figure~\ref{fig:task}), there are rich categorical data for these objects (a.k.a., static feature in Figure~\ref{fig:task}).
We use $\bm{x}^{u}_p$, $\bm{x}^{a}_n$, $\bm{x}^{i}_q$ to denote the feature of the $p$-th user, the $n$-th anchor, the $q$-th item respectively, and use $\bm{x}^{uj}_p$, $\bm{x}^{aj}_n$, $\bm{x}^{ij}_q$ to denote their $j$-th categorical features.
For convenience, we use the $1$-th categorical feature of each item to denote its category (e.g., shoes).
Namely, $\bm{x}^{i1}_q$ represents the $q$-th item's category.

\subsection{Related Work}
\minisection{Advanced Recommendation Tasks}
Classical item-based recommender systems \citep{koren2009matrix,cheng2016wide} aim at seeking the favorite item for a user according to user-item interactions.
Influenced by the evolution of mobile Internet and the development of deep learning techniques, more and more research in recommendation has shifted to inventing new recommendation tasks, which can be roughly categorized into two lines.
One line of literature \citep{hidasi2015session,ren2019lifelong,fan2019metapath,sun2018conversational} seeking to bring more convenience for users by modifying the recommendation task.
For example, \citet{hidasi2015session} introduces the session-based recommendation task where the recommender systems are only accessible to short session-based data instead of long histories.
Recently, \citet{fan2019metapath} proposed an intent recommendation to recommend an intent (i.e., query) to a user since typing words on mobile devices are much inconvenient than on desktop computers. 
The other line of work \citep{shi2018heterogeneous,chen2017attentive,van2013deep} investigating to include the side information associated with users and items in the recommendation to better capture user preference. 
For example, one promising way is to combine the structure information hidden in the sequence \citep{wu2017recurrent} or the graph \citep{shi2018heterogeneous} built based on user-item history in the recommendation.
Another direction is to model multimedia data such as image \citep{chen2017attentive} or audio \citep{van2013deep} related with recommendation. 
To the best of our knowledge, since the live broadcast recommendation is recently emerged and is developing rapidly, there is limited literature in this field, and our paper is proposed to fill this gap.

\minisection{Collaborative Filtering Methods}
Classical recommendation methods such as matrix factorization \citep{koren2009matrix} parameterize users and items as an embedding vector and conduct inner product between them to predict an interaction.
For further mining interactive information among features, FM \citep{rendle2010factorization} projects each feature into a low-dimensional vector and models feature interactions by the inner product.
As discussed in \citep{he2017neurala,hsieh2017collaborative}, although the inner product encourages user and item embeddings of an observed interaction close to each other, its natural linearity makes it insufficient to discover the complex correlations between users and items.
Influenced by the stunning success of deep learning, recent efforts \citep{he2017neurala,he2017neuralb,wu2016collaborative,tay2018latent,wang2019neural} focus on exploiting deep learning techniques to enhance the interaction function.
For instance, neural collaborative filtering models such as NeuMF \citep{he2017neurala} use non-linear functions to capture interactive patterns between users and items, translation based collaborative filtering models such as LRML \citep{tay2018latent} employ Euclidean distance metrics in modeling interaction.
DeepFM \citep{guo2017deepfm} incorporates an FM layer to replace the wide component in Wide \& Deep \citep{cheng2016wide}, PNN \citep{qu2016product} leverages a multi-layer perceptron (MLP) to model interaction of a product layer and recently proposed collaborative filtering methods working on structured data such as GraphHINGE \citep{jin2020learning} propose a new convolution-based interaction module on the heterogeneous graph.
The live broadcast recommendation scenarios, as mentioned above, are many complicated and heterogeneous situations.
Hence, we not only extend SVD++ \citep{koren2008factorization} to model correlations in our two-side architecture but propose a novel co-retrieval model collaborative filtering user's and anchor's relevant histories.

Our architecture design is also closely related to the two-side sequential networks.
For item recommendation task, there are recently emerged literature \citep{fu2021dual,wu2017recurrent,wu2019dual,qin2020sequential,wang2019neural} leveraging the context information from user and item sides to make the final prediction.
Besides the difference between the live broadcast recommendation and item recommendation tasks, these approaches either consider two-side information in an independent fashion \citep{wu2017recurrent,wu2019dual} or model two-side correlations among all the high-order neighbor users and items \citep{fu2021dual,qin2020sequential,wang2019neural} which is insufficient in filtering out noise and employing to long sequential data.
In contrast, we develop a novel interactive network with a co-retrieval mechanism to efficiently capture the key interactive patterns from two sides.

\section{The TWINS Model}

\subsection{Overview}
The basic idea of the TWINS is to design a two-side architecture to capture the rich context hidden in static and dynamic features in both user and anchor sides.
Figure~\ref{fig:overview} illustrates the overview of TWINS.
First, we use the PNN \citep{qu2016product} to model the correlations among static attributes for each user, anchor, and item, as shown in (a)-(d).
Second, we leverage the RNN \citep{hochreiter1997long} to capture the sequential dependencies hidden in the user's and anchor's dynamic histories, as shown in (b)(c). 
Third, we design interactive networks for mining the interactive patterns between user and anchor sides from item and anchor aspects, as shown in (e)(f).
Moreover, notice that the interaction operations, especially for interactive networks in item aspect, require the high computation cost; we propose a co-retrieval mechanism to select the relevant items from the whole user's and anchor's histories to save interaction computations, as shown in (h).
After that, we aggregate these interaction results accompanied with static features of users and anchors to predict the probability that a user will browse an anchor's broadcast room,  as shown in (i).
We introduce these steps in detail in the following subsections.

\subsection{Object Modeling}
In classical recommendation models, there are two main objects (i.e., users, items), while in the live broadcast recommendation scenario, there are three main objects (i.e., users, anchors, items).
As illustrated in Figure~\ref{fig:task}, for each pair of user and anchor, we have the static features (attributes), which are also called categorical data.
Notably, there are rich correlations among these features.
For example, the reason why the teen would be interested in the lipstick in a bright color should both rely on her age \textbf{AND} gender. 
As discussed in \citep{guo2017deepfm,jin2020efficient,qu2016product}, these ``AND'' operations can't solely be modeled by neural networks.
Hence, we introduce the PNN, whose  
output of the PNN for the $p$-th user can be defined as
\begin{equation}
\label{eqn:fm}
\bm{e}^u_p \coloneqq \mathtt{f}^u_\mathtt{PNN}(\bm{x}^{u}_p)= \bm{v}^u_p \odot \bm{x}^u_p + \sum^J_{j'=1}\sum^J_{j''=j'+1} (\bm{v}^u_{j'} \odot \bm{v}^u_{j''}) \bm{x}^{uj'}_p\cdot\bm{x}^{uj''}_p,
\end{equation}
where $\bm{v}^u_p$ and $\bm{v}^u_{j}$ are trainable latent vectors and $\odot$ is the element-wise product operator.
The first term is addition units showing the influence of (first-order) linear feature interactions, and the second term is element-wise product units representing the impact of (second-order) pair-wise feature interactions.

Consider that there are also static features for each anchor and item.
Analogously, we define $\bm{e}^a_n\coloneqq \mathtt{f}^a_\mathtt{PNN}(\bm{x}^{a}_n)$ as the output of the $n$-th anchor and $\bm{e}^i_q\coloneqq \mathtt{f}^i_\mathtt{PNN}(\bm{x}^{i}_q)$ as the output of the $q$-th item where $\mathtt{f}^a_\mathtt{PNN}(\cdot)$ and $\mathtt{f}^i_\mathtt{PNN}(\cdot)$ share the same formulation with $\mathtt{f}^u_\mathtt{PNN}(\cdot)$ but with different parameters.

Notice that besides the static features, the portrait of a user and the theme of an anchor are closely related to their dynamic histories, such as user's browsing items and anchor's broadcasting items, as illustrated in Figure~\ref{fig:overview}(b)(c).
A principal way to model these dynamic histories is to construct a sequential model such as the RNN model.
Let $\bm{h}^i_q$ denote the output of the $q$-th item, which can be calculated by
\begin{equation}
\label{eqn:rnn}
\bm{h}^i_q \coloneqq \mathtt{f}^i_\mathtt{RNN}(\bm{e}^i_q|\bm{b}^i_{q-1}),
\end{equation}
where $\mathtt{f}^i_\mathtt{RNN}(\cdot)$ is the RNN cell 
and $\bm{b}^i_{q-1}$ is the hidden vector computed from the last RNN cell.
In our paper, we implement the RNN cell as a standard LSTM unit \citep{hochreiter1997long}. 
As the major objects for browsing and broadcasting are items, we only build the RNN model for sequences of items.

\subsection{Interactive Network}
By encoding the static and dynamic features in triple objects, we obtain the embedding vectors of the $p$-th user (i.e., $\bm{e}^u_p$), the $q$-th item (i.e., $\bm{h}^i_q$), and $n$-th anchor (i.e., $\bm{e}^a_n$), as shown in Figure~\ref{fig:overview}.
We then consider mining the interactive patterns by the ``AND'' operation.
The motivation behind this is straightforward.
Take Figure~\ref{fig:overview} as an instance.
The teen $u_1$ enters the broadcasting room because the host anchor $a_1$ sells her favored items \textbf{AND} the anchor $a_1$ shares the similar interest with her favored anchors.
Thus, we model these interactive patterns in two aspects, namely item and anchor aspects.

\minisection{Item Aspect}
For item aspect, as illustrated in Figure~\ref{fig:overview}(e), TWINS captures the interactive patterns by measuring the similarities between user and anchor together with their related items.
A principal way is to follow the basic idea of SVD++ \citep{koren2008factorization} model, and
then the interaction similarity of the $p$-th user $u_p$ and the $n$-th anchor $a_n$ can be formulated as 
\begin{equation}
\label{eqn:svd}
y^i_{pn} = (\bm{e}^u_p + \sum_{q'\in \mathcal{H}^{ui}_{p}}\lambda_{pq'}\bm{h}^i_{q'})^\top \cdot (\bm{e}^a_n+\sum_{q''\in\mathcal{H}^{ai}_n}\beta_{nq''}\bm{h}^i_{q''}).
\end{equation}
Clearly by assigning $\lambda_{pq'}$ and $\beta_{nq''}$ as $1/\sqrt{|\mathcal{H}^{ui}_{p}|}$ and $0$ separately, we can exactly recover using the classical SVD++ model to measure the similarity between $u_p$ and $a_n$.
Notably, as users browsing the same items are normally diversified, it's non-trivial to capture the useful information from abundant context information of these users.
Hence, the classical SVD++ model, originally proposed for the similarity between users and items, doesn't involve this information (i.e., $\beta_{nq''}=0$).
Instead, as shown in the right part of Eq.~(\ref{eqn:svd}), we use the broadcast items to enrich the representation of $a_n$, which is much clean and informative.

As discussed in \citep{wang2019neural,qu2019end,jin2020efficient}, many existing methods 
(including the interactive network built following Eq.~(\ref{eqn:svd}))
suffer from the ``early summarization'' issue, as these approaches, when extending to similarity measurement between users and anchors, usually compress user-related and anchor-related items into single user/anchor embeddings before the final prediction.
In this case, only two objects are activated, yet other related objects (e.g., items) and their correlations are mixed and relayed.
We argue that these rich correlations (i.e., interactive patterns) are essential in the recommendation.
Taking Figure~\ref{fig:overview}(e) as an instance, a system is considering to recommend an anchor (e.g., $a_1$) to a user (e.g., $u_1$).
Suppose that $u_1$'s personal interest mainly lies in shoes,
then the similarity between ($i_1$ and $i_5$), ($i_3$ and $i_5$) should be emphasized.
Therefore, we propose a bi-attention network to better capture these interactive patterns, which can be formulated as follows:
\begin{equation}
\label{eqn:biattentioni}
\begin{aligned}
\alpha^i_{pq'nq''} &= \mathtt{f}^i_\mathtt{softmax}(\bm{w}_{pq'nq''}^{i\top}[\bm{e}^u_p,\bm{h}^i_{q'}, \bm{e}^a_n,\bm{h}^i_{q''}]+b^i_{pq'nq''}),\\
\bm{y}^i_{pn} &= \sum_{q'\in \mathcal{H}^{ui}_{p}}\sum_{q''\in\mathcal{H}^{ai}_n} \alpha_{pq'nq''}  (\bm{h}^i_{q''} \odot \bm{h}^i_{q''}),
\end{aligned}
\end{equation}
where $[\cdot,\cdot]$ denotes a concatenation operator. and $\mathtt{f}^i_\mathtt{softmax}(\cdot)$ denotes a softmax function.
Comparing to Eq.~(\ref{eqn:svd}), Eq.~(\ref{eqn:biattentioni}) takes both user- and anchor-side items to generate differentiable weights distinctive to different interaction terms.


\minisection{Anchor Aspect}
For the anchor aspect, as shown in Figure~\ref{fig:overview}(f), TWINS aims to formulate the similarities between the user along with her browsed anchors and target anchor.
Sharing the same motivation with an interactive network of item aspect, we formulate the interaction operation as follows:
\begin{equation}
\label{eqn:biattentiona}
\begin{aligned}
\alpha^a_{pn'n} &= \mathtt{f}^a_\mathtt{softmax}(\bm{w}_{pn'nn''}^{i\top}[\bm{e}^u_p,\bm{e}^a_{n'}, \bm{e}^a_n]+b^a_{pn'n}),\\
\bm{y}^a_{pn} &= \sum_{n'\in \mathcal{H}^{ua}_{p}} \alpha_{pn'n} (\bm{e}^a_{n'} \odot \bm{e}^a_{n}),
\end{aligned}
\end{equation}
where $\mathtt{f}^a_\mathtt{softmax}(\cdot)$ denotes a softmax function with different weight from $\mathtt{f}^i_\mathtt{softmax}(\cdot)$.

\subsection{Co-Retrieval Mechanism}
Notably, comparing Eq.~(\ref{eqn:biattentioni}) to Eq.~(\ref{eqn:biattentiona}), one can easily see that interactive networks of item aspect require to compute the similarity among $|\mathcal{H}^{ui}_p|\times|\mathcal{H}^{ai}_n|$ operations for each user-anchor pair $(u_p,a_n)$ which is much more time-consuming than that of anchor aspect whose computation costs lie in $|\mathcal{H}^{ua}_p|$ operations.
Therefore, the former one blocks TWINS from working in the real-world industrial scenario, especially with long sequential data \citep{pi2020search}.

In order to effectively implement the interactive network of item aspect, we introduce a novel co-retrieval mechanism, whose basic idea is to find a subset of user's and anchor's related items to feed in the network instead of using the whole data.

Inspired by recently merged search-based methods \citep{pi2020search,qin2020user}, 
we design a hard-search co-retrieval model without any parametric, where
only items belongs to the common categories of user and anchor sides will be selected as the candidate items to feed into the interactive network.
Formally, we first construct a set of categories for user and anchor sides respectively, namely $\mathcal{C}^u_p=\{\bm{x}^{i1}_{q'}|i_{q'}\in \mathcal{H}^{ui}_p\}$ and $\mathcal{C}^a_n=\{\bm{x}^{i1}_{q''}|i_{q''}\in \mathcal{H}^{ai}_n\}$.
We then compute a set of the common categories as $\mathcal{C}^{ua}_{pn}=\mathcal{C}^u_p \cap \mathcal{C}^a_n$.
We establish a retrieved set of $\mathcal{H}^{ui}_p$ and $\mathcal{H}^{ai}_n$ in Eq.~(\ref{eqn:biattentioni}) by following
\begin{equation}
\label{eqn:coretrieval}
\begin{aligned}
\widehat{\mathcal{H}}^{ui}_p &= \{i_{q'}|i_{q'}\in \mathcal{H}^{ui}_p \text{ and } \bm{x}^{i1}_{q'} \in\mathcal{C}^{ua}_{pn}\},\\
\widehat{\mathcal{H}}^{ai}_n &= \{i_{q''}|i_{q''}\in \mathcal{H}^{ai}_n \text{ and } \bm{x}^{i1}_{q''} \in\mathcal{C}^{ua}_{pn}\}.
\end{aligned}
\end{equation}
Clearly, $\widehat{\mathcal{H}}^{ui}_p$ and $\widehat{\mathcal{H}}^{ai}_n$ are subsets of $\mathcal{H}^{ui}_p$ and $\mathcal{H}^{ai}_n$ respectively.
One can directly replace $\mathcal{H}^{ui}_p$, $\mathcal{H}^{ai}_n$ by $\widehat{\mathcal{H}}^{ui}_p$, $\widehat{\mathcal{H}}^{ai}_n$ in Eq.~(\ref{eqn:biattentioni}) to save computations.

\subsection{Optimization Objective}
After primitively modeling each object and further interactive pattern mining, for each user-anchor pair (e.g., ($u_p,a_n$)), we can obtain the similarity based on their embedding vector namely $\bm{y}^e_{pn}= \bm{e}^u_p$ $\odot$ $\bm{e}^a_n$.
As we have already obtained item aspect interaction result $\bm{y}^i_{pn}$ and anchor aspect interaction result $\bm{y}^a_{pn}$, we further aggregate them together to produce the final similarly by combining a sigmoid function with a MLP layer over the concatenation of these embeddings as
\begin{equation}
\label{eqn:y}    
\widehat{y}_{pn} = \text{sigmoid}(\mathtt{f}_\mathtt{MLP}([\bm{y}^e_{pn},\bm{y}^i_{pn},\bm{y}^a_{pn}])).
\end{equation}
We then use the log loss as the objective:
\begin{equation}
\label{eqn:loss}
\mathcal{L} = - \sum_{(u_p,a_n)\in\mathcal{D}}(y_{pn}\log \widehat{y}_{pn}+(1-y_{pn})\log(1-\widehat{y}_{pn})),
\end{equation}
where $\mathcal{D}=(\mathcal{U},\mathcal{A})$ denotes the dataset and $y_{pn}$ is the label of each user-anchor instance.

We provide the learning algorithm of TWINS in Algorithm~\ref{algo:framework}.
We also provide the corresponding analysis of TWINS in Appendix~\ref{app:algorithm}.

\begin{algorithm}[t]
	\caption{TWINS}
	\label{algo:framework}
	\begin{algorithmic}[1]
		\REQUIRE
		dataset $\mathcal{D}=(\mathcal{U}, \mathcal{A})$ with historical data $\mathcal{H}^u$, $\mathcal{H}^a$;
		\ENSURE
		TWINS recommender with parameter $\theta$
		\vspace{1mm}
		\STATE Initialize all parameters.
		\REPEAT
		\STATE Randomly sample a batch $\mathcal{B}$ from $\mathcal{D}$
		\FOR {each data instance $(u_p, a_n)$ in $\mathcal{B}$}
		\STATE Calculate embedding vectors for all related user, anchors, items using static features via FM model as Eq.~(\ref{eqn:fm}).
		\STATE Compute embedding vectors for all sequential items using dynamic features via RNN model as Eq.~(\ref{eqn:rnn}).
		\STATE Obtain item aspect similarity $\bm{y}^i_{pn}$ using Eq.~(\ref{eqn:biattentioni}).
		\STATE Obtain anchor aspect similarity $\bm{y}^a_{pn}$ using Eq.~(\ref{eqn:biattentiona}).
		\ENDFOR
		\STATE Compute $\mathcal{L}$ and update $\theta$ by minimizing Eq.~(\ref{eqn:loss}).
		\UNTIL convergence
	\end{algorithmic}
\end{algorithm}


\section{Offline Experiments}

\subsection{Dataset and Experimental Flow}
We conduct offline experiments on four real-world datasets, namely Yelp business dataset\footnote{\url{https://www.yelp.com/dataset/documentation/main}}, Trust statement dataset\footnote{\url{http://www.trustlet.org/downloaded_epinions.html}}, Aminer citation dataset\footnote{\url{https://www.aminer.cn/citation}}, Diantao live broadcast dataset, where the first three are public benchmark datasets and the last one is created by our own.
We provide detailed description of the last dataset as follows, and offer the description of others in Appendix~\ref{app:dataset}.
\begin{itemize}[topsep = 3pt,leftmargin =5pt]
	\item \textbf{Diantao Live Broadcast Recommendation dataset} is collected from the user interaction logs of Diantao App.
	It contains more than 1.46 billion logs of over 10 million users' browsing histories with 90 thousand anchors.
	Features of the user include age, gender, city, etc., and features of the document include title, time, etc.   
	In each query, we regard the documents whose playtime are more than $3$s as the clicked ones.
	\end{itemize}
Please refer to Appendix~\ref{app:config} for detailed experimental configuration.

\subsection{Baseline and Evaluation Metric}
We make comprehensive comparisons between our model and 9 representative baseline methods, introduced as follows. 
\begin{itemize}[topsep = 3pt,leftmargin =5pt]
\item \textbf{FM} \citep{rendle2010factorization} is the factorization machine that uses the linear projection and inner product of features to measure the user-item similarity. 
\item \textbf{NeuMF} \citep{he2017neurala} is a generalized model consisting of a matrix factorization (MF) and a MLP component.
\item \textbf{DeepFM} \citep{guo2017deepfm} is a generalized model consisting of a FM as a wide component and a MLP as a deep component. 
\item \textbf{PNN} \citep{qu2016product} is the product-based neural network consisting of a embedding layer and a product layer to capture interactive patterns.
\item \textbf{LSTM} \citep{hochreiter1997long} is the long short term memory network widely used to model sequential data.
\item \textbf{NARM} \citep{wu2017recurrent} is a sequential recommendation model, which uses attention mechanism to capture the influence of user behaviors.
\item \textbf{ESMM} \citep{li2017neural} is a multi-objective model which applies a feature representation transfer learning strategy on user behaviors.  
\item \textbf{DIN} \citep{zhou2018deep} designs a local activation unit to adaptively learn the representation of user interests from historical behaviors.
\item \textbf{DIEN} \citep{zhou2019deep} builds an interest extractor layer based on DIN to capture temporal interests from historical behavior sequence.
\end{itemize}
Note that as all these methods are originally proposed particularly for classical item-based recommendation tasks definitively different from the live broadcast recommendation task, thus we introduce two versions of implementation.
Taking LSTM as an instance, we use the model for the historical sequences of user browsed anchors (denoted as LSTM$^-$).
Also, we can first use LSTM to model the historical sequences of user browsed anchors, anchor broadcast items, user browsed items, and then fuse this information via a MLP layer with a sigmoid function to generate the final prediction (denoted as LSTM). 
For those tabular recommendation models such as FM, we apply the model for binary interactions between users and anchors (denoted as FM).

In order to further investigate the effect from each component of TWINS, we design the following three variants:
\begin{itemize}[topsep = 3pt,leftmargin =5pt]
\item \textbf{TWINS} is our model without co-retrieval mechanism.
\item \textbf{TWINS}$^-_i$ is a variant of TWINS, applying the original model without the interactive network from item aspect.
\item \textbf{TWINS}$^-_a$ is a variant of TWINS, applying the original model without the interactive network from anchor aspect.
\item \textbf{TWINS}$^+_\text{co}$ is a variant of TWINS using co-retrieval mechanism.
\end{itemize}
To evaluate the above methods, we choose Area user the ROC Curve (AUC), Accuracy (ACC), LogLoss as evaluation measurements.
The threshold of ACC of all the datasets is set as 0.5.

\begin{figure}[t]
	\centering
	\includegraphics[width=1.0\linewidth]{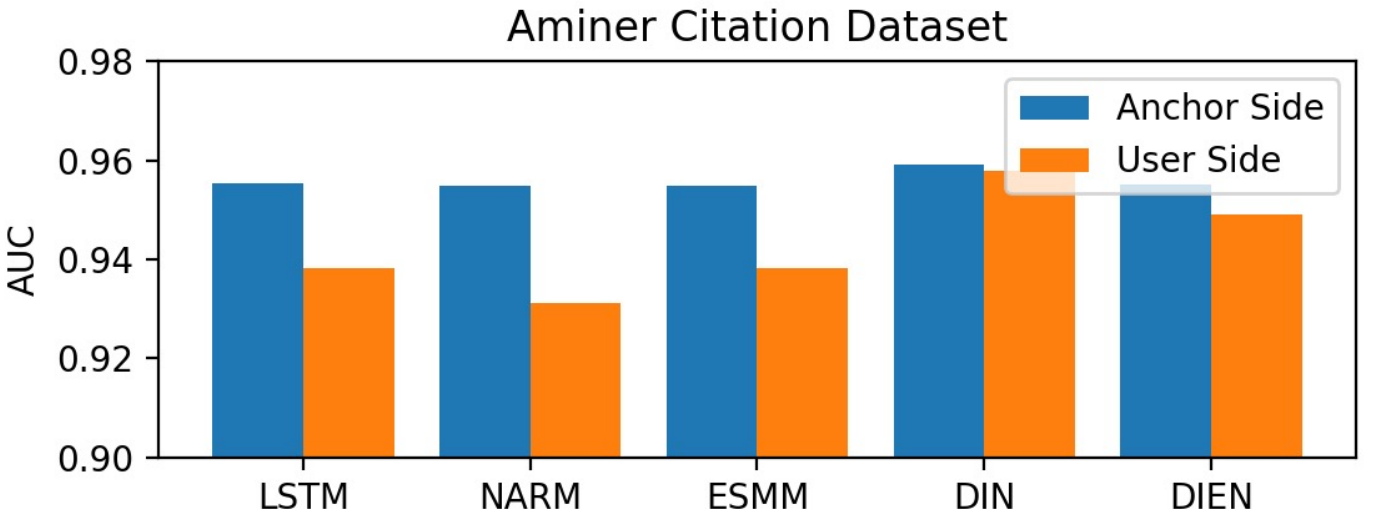}
	\vspace{-7mm}
	\caption{
		Comparisons of performance of baselines taking the information of anchor side or user side as the input on Aminer citation dataset.
	}
	\label{fig:side}
	\vspace{-3mm}
\end{figure}

\begin{table*}[t]
	\centering
	\caption{Comparison of different models on four industrial datasets. Results of Click-Through Rate (CTR) in term of AUC, ACC, LogLoss are reported. 
	* indicates $p < 0.001$ in significance tests compared to the best baseline.
	}
	\vspace{-2mm}
	\resizebox{1.00\textwidth}{!}{
		\begin{tabular}{@{\extracolsep{4pt}}ccccccccccccc}
		\toprule
			\multirow{2}{*}{Methods} & \multicolumn{3}{c}{Yelp Business Dataset} & \multicolumn{3}{c}{Trust Statement Dataset} & \multicolumn{3}{c}{Aminer Citation Dataset} & \multicolumn{3}{c}{Diantao Live Broadcast Dataset} \\
			\cmidrule{2-4}
			\cmidrule{5-7}
			\cmidrule{8-10}
			\cmidrule{11-13}
			{} & LogLoss & ACC & AUC & LogLoss & ACC & AUC & LogLoss & ACC & AUC & LogLoss & ACC & AUC\\
			\midrule
			FM & 
			0.6677 & 0.5945 & 0.6233 & 
			0.6188 & 0.7003 & 0.7574 & 
			0.6510 & 0.6430 & 0.7071 & 
			0.6541 & 0.6732 & 0.6896 \\
			\midrule
			NeuMF & 
			0.4895 & 0.7759 & 0.8424 & 
			0.4814 & 0.7835 & 0.8440 & 
			0.4797 & 0.7882 & 0.8542 & 
			0.5942 & 0.7124 & 0.7229 \\
			\midrule
			DeepFM &
			0.4594 & 0.7925 & 0.8658 & 
			0.4545 & 0.8054 & 0.8756 & 
			0.4410 & 0.8049 & 0.8826 & 
			0.5832 & 0.7231 & 0.7345 \\
			\midrule
			PNN & 
			0.4581 & 0.7931 & 0.8668 &
			0.4452 & 0.8144 & 0.8838 &
			0.3932 & 0.8789 & 0.9399 &
			0.5432 & 0.7367 & 0.7578 \\
			\midrule
			DIN$^-$ &
			0.3256 & 0.8731 & 0.9420 &
			0.4164 & 0.8457 & 0.9154 &
			0.1847 & 0.9423 & 0.9799 &
			0.5231 & 0.7564 & 0.7790 \\
			\midrule
			DIN &
			0.3156 & 0.8782 & 0.9451 &
			0.3771 & 0.8394 & 0.9114 &
			0.1212 & 0.9581 & 0.9892 &
			0.5100 & 0.7784 & 0.7995 \\
			\midrule
			LSTM$^-$ &
			0.3236 & 0.8736 & 0.9433 &
			0.3931 & 0.8271 & 0.9012 &
			0.2218 & 0.9204 & 0.9708 &
			0.5334 & 0.7529 & 0.7602 \\
			\midrule
			LSTM &
			0.3204 & 0.8783 & 0.9445 &
			0.3854 & 0.8325 & 0.9051 &
			0.1214 & 0.9575 & 0.9891 &
			0.5321 & 0.7602 & 0.7789 \\
			\midrule
			NARM$^-$ &
			0.3132 & 0.8811 & 0.9463 &
			0.3916 & 0.8306 & 0.9037 &
			0.2156 & 0.9216 & 0.9720 &
			0.5421 & 0.7432 & 0.7667 \\
			\midrule
			NARM &
			0.3137 & 0.8808 & 0.9463 &
			0.3839 & 0.8339 & 0.9060 &
			0.1200 & 0.9580 & 0.9893 &
			0.5233 & 0.7756 & 0.7953 \\
			\midrule
			ESMM$^-$ &
			0.3224 & 0.8722 & 0.9414 &
			0.3959 & 0.8268 & 0.9008 &
			0.2402 & 0.9110 & 0.9655 &
			0.5334 & 0.7456 & 0.7698 \\
			\midrule
			ESMM &
			0.3150 & 0.8776 & 0.9448 &
			0.4035 & 0.8262 & 0.8991 &
			0.1241 & 0.9564 & 0.9887 &
			0.5175 & 0.7753 & 0.7985 \\
			\midrule
			DIEN$^-$ &
			0.3198 & 0.8723 & 0.9401 &
			0.4018 & 0.8388 & 0.9091 &
			0.2254 & 0.9183 & 0.9698 &
			0.5252 & 0.7657 & 0.7854 \\
			\midrule
			DIEN &
			0.3291 & 0.8646 & 0.9395 &
			0.3911 & 0.8308 & 0.9022 &
			0.1242 & 0.9563 & 0.9887 &
			0.5145 & 0.7843 & 0.8046 \\
			\midrule
			TWINS$^-_i$ &
			0.2746 & 0.8879 & 0.9538 &
			0.3685 & 0.8452 & 0.9174 &
			0.1246 & 0.9616 & 0.9893 &
			0.5012 & 0.7896 & 0.8010 \\
			\midrule
			TWINS$^-_a$ &
			0.2608 & 0.8948 & 0.9583 &
			0.3660 & 0.8446 & 0.9170 &
			0.1233 & 0.9622 & 0.9895 &
			0.4977 & 0.7920 & 0.8024 \\
			\midrule
			\textbf{TWINS} &
			\textbf{0.2603}$^*$ & \textbf{0.9120}$^*$ & \textbf{0.9659}$^*$ &
			\textbf{0.3528}$^*$ & \textbf{0.8501}$^*$ & \textbf{0.9235}$^*$ &
			\textbf{0.1081}$^*$ & \textbf{0.9631}$^*$ & \textbf{0.9913}$^*$ &
			\textbf{0.4855}$^*$ & \textbf{0.7934}$^*$ & \textbf{0.8187}$^*$ \\
			\midrule
			\textbf{TWINS}$^+_\text{co}$ & 
			\textbf{0.2596}$^*$ & \textbf{0.8962}$^*$ & \textbf{0.9593}$^*$ &
			\textbf{0.3579}$^*$ & \textbf{0.8458}$^*$ & \textbf{0.9194}$^*$ &
			\textbf{0.1170}$^*$ & \textbf{0.9612}$^*$ & \textbf{0.9903}$^*$ &
			\textbf{0.4731}$^*$ & \textbf{0.8045}$^*$ & \textbf{0.8205}$^*$ \\ 
			\bottomrule
		\end{tabular}
	}
	\label{tab:res}
	\vspace{-1mm}
\end{table*}

\subsection{Performance Evaluation}
\minisection{Overall Performance}
Table~\ref{tab:res} summarizes the results.
The major findings from our experiments are summarized as follows.
\begin{itemize}[topsep = 3pt,leftmargin =5pt]
\item Compared to the version of only using user browsed anchors (denoted as X$^-$ and X can be DIN, LSTM, NARM, ESMM, DIEN), in most cases, X achieves better performance, which verifies to further include user browsed items and anchor browsed items as the input.
One also observe in some cases, X$^-$ obtains better performance, which may be explained as a simple aggregation operation (e.g., concatenation) that can not fully use this information, sometimes even bringing the noise.
\item Our model outperforms all these baseline methods, including widely adopted industrial recommendation methods (e.g., DeepFM, ESMM, DIN, DIEN), interaction models (e.g., FM, PNN), and sequential models (e.g., NARM, LSTM).
As the inputs are the same, these results would indicate the superiority of developing interactive networks based on the two-side architecture. 
\item With the comparison between LSTM to other baseline methods, we see that LSTM can consistently achieve comparable or even better performance than interaction models (i.e., FM, NeuMF, DeepFM, PNN), which verifies the necessity of mining the sequential patterns of users and anchors.
\end{itemize}

\minisection{Impact of Interaction Networks}
From comparisons between TWINS and TWINS$^-_i$, TWINS and TWINS$^-_a$, TWINS consistently achieves better performance than TWINS$^-_i$ and TWINS$^-_a$ in all the datasets.
One explanation is that our interactive networks are able to provide interactive (i.e., ``AND'') patterns, which can not be solely modeled by employing a neural network.
By comparing TWINS$^-_i$ to TWINS$^-_a$, TWINS$^-_a$ usually can gain better performance.
One possible reason for this is that in the live broadcast recommendation system, the similarities between users and anchors mainly depend on their browsed and broadcast items.
Namely, the reason that a user watches an anchor mainly lies in that the anchor is selling some items that she is interested in. 

\begin{figure}[t]
	\centering
	\includegraphics[width=1.0\linewidth]{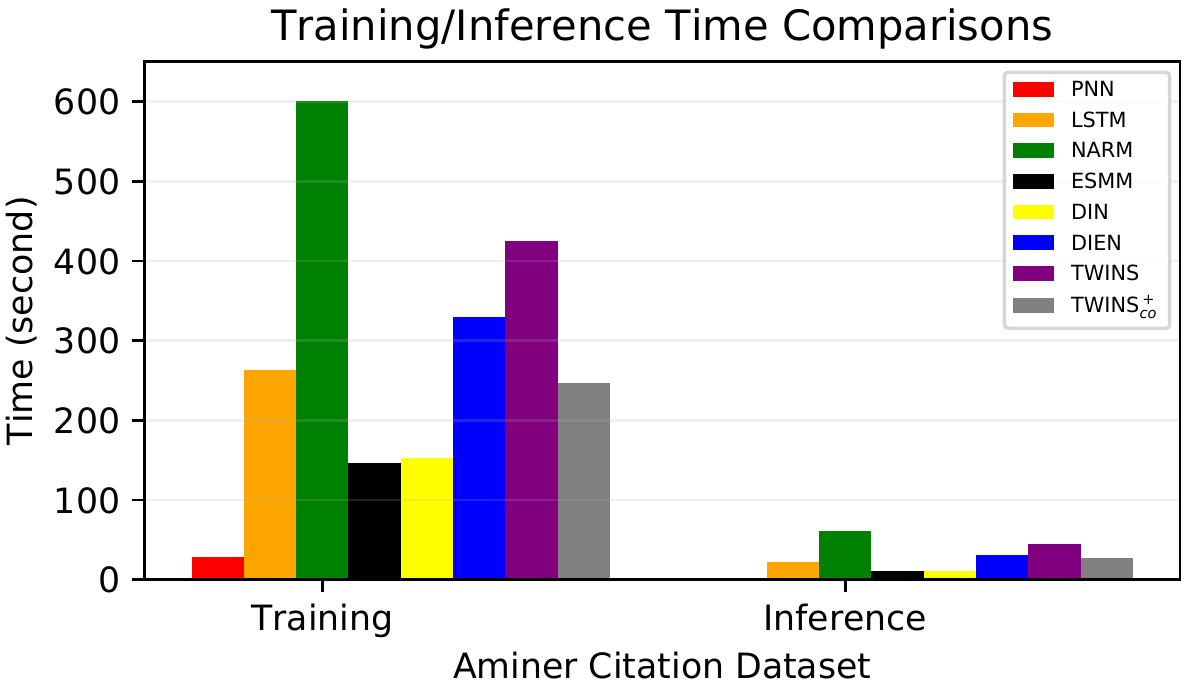}
	\vspace{-7mm}
	\caption{
		Training/inference time comparisons of TWINS and TWINS$_\text{co}$ against baselines on Aminer citation dataset.
	}
	\label{fig:time}
	\vspace{-4mm}
\end{figure}

\minisection{Impact of Co-Retrieval Mechanism}
Comparing TWINS$^+_\text{co}$ to TWINS, we can observe that TWINS$^+_\text{co}$ can achieve a comparable, or even better, result in these datasets.
This result is consistent with the result of the hard-search reported in \citep{pi2020search}, both of which reveal that category information plays a vital role in selecting relevant items.
We then further report their training and inference time in Figure~\ref{fig:time} to verify that TWINS with the proposed co-retrieval mechanism is more efficient and thus could deal with long-sequential data.

\minisection{Impact of Two-Side Information}
As introduced in Section~\ref{sec:problem}, there are three sequential data in user and anchor sides, namely a sequence of user browsed anchors denoted as $\mathcal{H}^{ua}$, a sequence of user browsed items denoted as $\mathcal{H}^{ui}$ and a sequence of anchor broadcast items denoted as $\mathcal{H}^{ai}$, where the first two sequences are of user side and the last one sequence is of anchor side.
Since the main difference between live broadcast recommendation and item recommendation is that the former one requires us to take the information of both user and anchor sides into consideration, while the latter one is designed to model the information from one sequential data of one side (usually $\mathcal{H}^{ua}$).
Results in Table~\ref{tab:res} shows the results of using $\mathcal{H}^{ua}$ (denoted as X$^-$) and using all these sequences fused by a concatenation operation (denoted as X).
We further investigate the performance gain of X$^-$ by adding $\mathcal{H}^{ui}$ (denoted as \texttt{User} \texttt{Side}) or $\mathcal{H}^{ai}$ (denoted as \texttt{Anchor} \texttt{Side}) into the input.
From Figure~\ref{fig:side}, we see that $\mathcal{H}^{ai}$ is more useful for X$^-$ than $\mathcal{H}^{ui}$.
One explanation is that we already have $\mathcal{H}^{ua}$ as the information on the user side and no the information on anchor side.
Hence, $\mathcal{H}^{ai}$ can offer more important information than $\mathcal{H}^{ui}$.

\minisection{Complexity Analysis}
We investigate the time complexity of TWINS and TWINS$_\text{co}$ against baseline methods such as PNN, LSTM, NARM, ESMM, DIN, DIEN, and report the training and inference times for one round of the whole data.
As Figure~\ref{fig:time} depicts, sequential methods (e.g., NARM, DIEN) are less efficient than other methods (e.g., PNN).
Also, we can see that TWINS$^+_\text{co}$ is more effective, as it can reduce the computation costs of interactive networks.
One can also use the co-retrieval mechanism in object modeling, where only retrieved items are fed into the RNN model instead of the whole set of items, to reduce the computation costs from the RNN model.

\section{Online Experiments}
\subsection{Experimental Flow}
In order to verify the effectiveness of TWINS$^+_\text{co}$ in real-world live broadcast recommendation applications, we deploy our method in Diantao App, a main-stream live broadcast App sharing all the anchors with Taobao e-commerce platform, which has tens of millions of daily active users who create hundreds of millions of user logs every day in the form of implicit feedbacks such as click, watch behaviors.
For simplicity, we use TWINS to denote our method and use TWINS$^+_\text{co}$ as the implementation.
We develop two kinds of techniques to light the current TWINS model and develop an effective data structure, as shown in Figure~\ref{fig:deploy}.
We further introduce their details along with our hands-on experience of implementing TWINS in Alibaba in Appendix~\ref{app:deploy}. 

\begin{table}[h]
	\centering
	\caption{Improvement of TWINS against current production method on real-world recommendation scenarios.}
	\vspace{-2mm}
	\resizebox{0.80\linewidth}{!}{
		\begin{tabular}{@{\extracolsep{4pt}}cccccccccc}
			\toprule
			Recommender & ACTR & UCTR & UCVR\\
			\midrule
			\textbf{TWINS} & 
			\textbf{8.11\%} & \textbf{2.01\%} & \textbf{3.52\%}\\
			\bottomrule
		\end{tabular}
	}
	\label{tab:deploy}
\end{table}

For the online experiment, we conduct A/B testing comparing the proposed model TWINS with the current production method.
The whole experiment lasts a week, from September 25, 2021 to October 2, 2021.
During A/B testing, 5\% of the users are presented with the recommendation by the current production method, while 5\% of the users are presented with the recommendation by TWINS.

\subsection{Performance Evaluation}
We examine the online performance using three metrics.
The first one is to measure the CTR performance from the anchor aspect, which is called ACTR metric defined as $\text{ACTR}=\frac{\# \text{clicks on anchors}}{\# \text{impressions on anchors}}$ where \#clicks on anchors and  \#impressions on anchors are the number of clicks and impressions on all the anchors. 
The second one is to measure the CTR performance from the user aspect, which is called UCTR metric defined as $\text{UCTR}=\frac{\# \text{clicks on users}}{\# \text{impressions on users}}$ where \#clicks on users is the number of users that have performed click behaviors, and \#impressions on users is the total number of users.
The third one is to measure the CVR performance from the user aspect, which is called UCVR metric defined as $\text{UCTR}=\frac{\# \text{conversions on users}}{\# \text{impressions on users}}$ where \#conversions on users is the number of users that have performed conversion behaviors, and \#impressions on users is the total number of users.
We report the average results in Table~\ref{tab:deploy}.
One can notice that TWINS consistently achieves better performance in terms of all the metrics.

\subsection{Case Study}
\begin{figure}[t]
	\centering
	\includegraphics[width=1.0\linewidth]{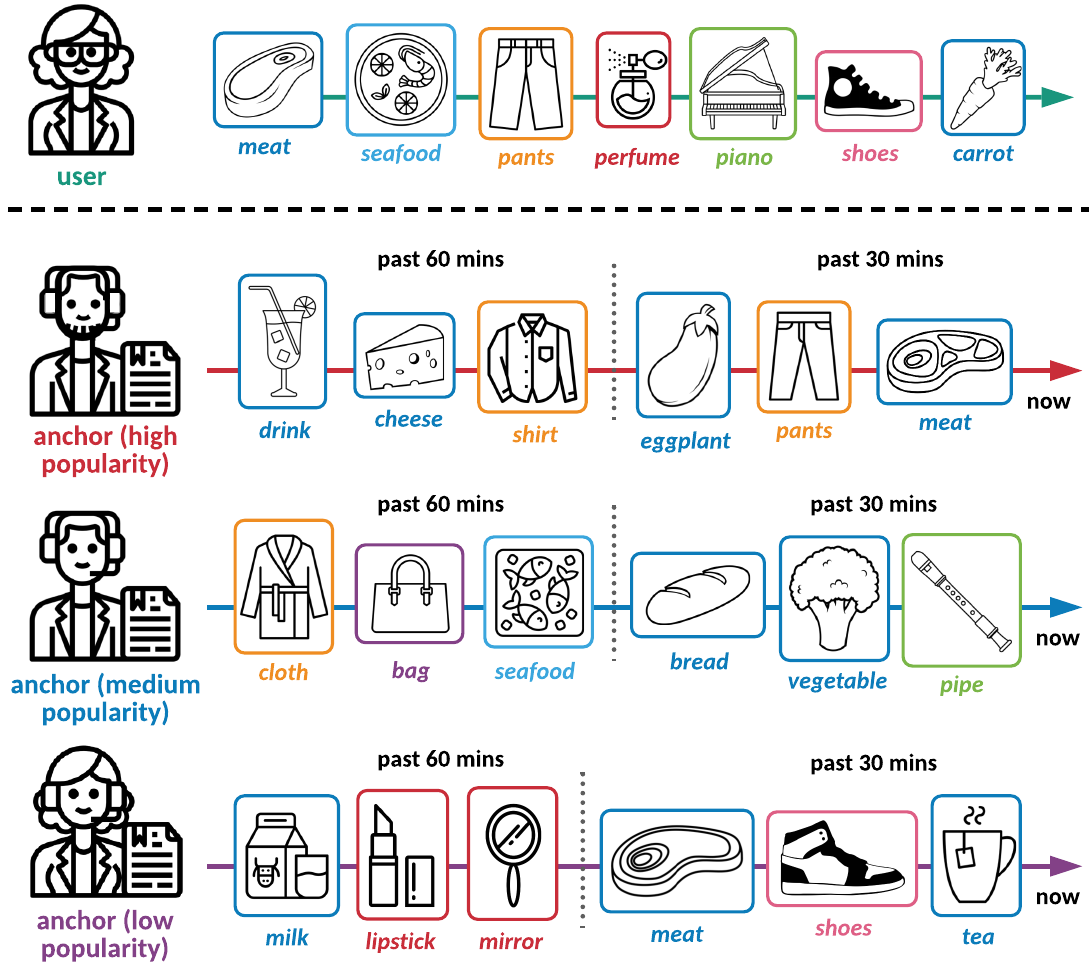}
	\vspace{-6mm}
	\caption{
		Illustration of the interaction patterns of TWINS in the case of predicting the relevance between a user and different anchors with the different popularity. 
		Each pair of Items with similar color demonstrates the high interaction weights (i.e., correlations) between two items.
	}
	\label{fig:case}
	\vspace{-2mm}
\end{figure}

Finally, we conduct case study to reveal the inner structure of TWINS on Diantao App.
Figure~\ref{fig:case} illustrates the interaction patterns between each pair of items in user and anchor sides, where the ones with similar colors means the high interaction weights.
As expected, we can see that these interaction weights can well reflect the corresponding correlations.
For example, clothes including pants and shirt have the same color (i.e., yellow), and have the similar color with cosmetics containing perfume and lipstick (i.e., red).
Based on them, TWINS can recommend appropriate anchors to the user.
We note that as shown in Figure~\ref{fig:case}, the recommended anchors can be in range from high popularity to low popularity.
We argue that it is quite meaningful in the practice nowadays where the top popular anchors can usually attach most users' attentions, which is similar to the popularity bias \citep{chen2020bias} in traditional recommendation task. 
Therefore, the proposed method can simultaneously improve the performance of the recommendation while mitigating the bias issue (i.e., not always recommending the anchors with the high popularity to different users).

\section{Conclusion and Future Work}
In this paper, we investigate a recently emerged live broadcast recommendation and propose a novel two-side framework named TWINS, where we design interactive networks from item and anchor aspects to capture the rich interactive patterns in user and anchor sides. 
In addition, we also develop a co-retrieval mechanism to reduce the high computation costs of the interactive network from the item aspect.
For future work, it would be interesting to combine TWINS with multi-tasking learning techniques to effectively use user various behaviors (e.g., click, like, comment).

\minisection{Acknowledgments} 
This work was supported by Alibaba Group through Alibaba Research Intern Program.
The Shanghai Jiao Tong University Team is supported by Shanghai Municipal Science and Technology Major Project (2021SHZDZX0102) and National Natural Science Foundation of China (62076161, 62177033).
We would also like to thank Wu Wen Jun Honorary Doctoral Scholarship from AI Institute, Shanghai Jiao Tong University.

\clearpage
\bibliographystyle{ACM-Reference-Format}
\bibliography{twins}
\clearpage
\appendix
\section{Model Analysis of TWINS}
\label{app:algorithm}

The learning algorithm of TWINS is given in Algorithm~\ref{algo:framework}.
As classical item recommendation methods \citep{koren2009matrix,qu2016product} often use the inner-product of user and item embedding vectors (i.e., $\bm{y}_{pn}^e$) to measure their similarity, we further clarify our motivations to involve interaction results from both item and anchor aspects.
Solely using $\bm{y}_{pn}^e$ for learning the model may be sufficient to fit the true conditional probability $P(y_{pn}|(u_p,a_n), \mathcal{H}_p^u, \mathcal{H}_n^a)$, if we are accessible to the labels of all the possible user-anchor pairs.
However, the limited user observations over anchors in practice would lead the limited performance.
Concretely, we can divide those unobserved samples into two parts, namely, unobserved positive and unobserved negative samples.
The former class refers to samples where the users would show positive feedbacks (e.g., click) if browsing the anchors, while the latter class refers to samples where the users would show negative feedbacks (e.g., not click) if browsing the anchors.
As under most circumstances, there is no auxiliary information to distinguish these two classes; all the unobserved samples are often directly treated as negative samples, which indeed provides wrong supervisions for learning the model.

\begin{figure}[t]
	\centering
	\includegraphics[width=1.0\linewidth]{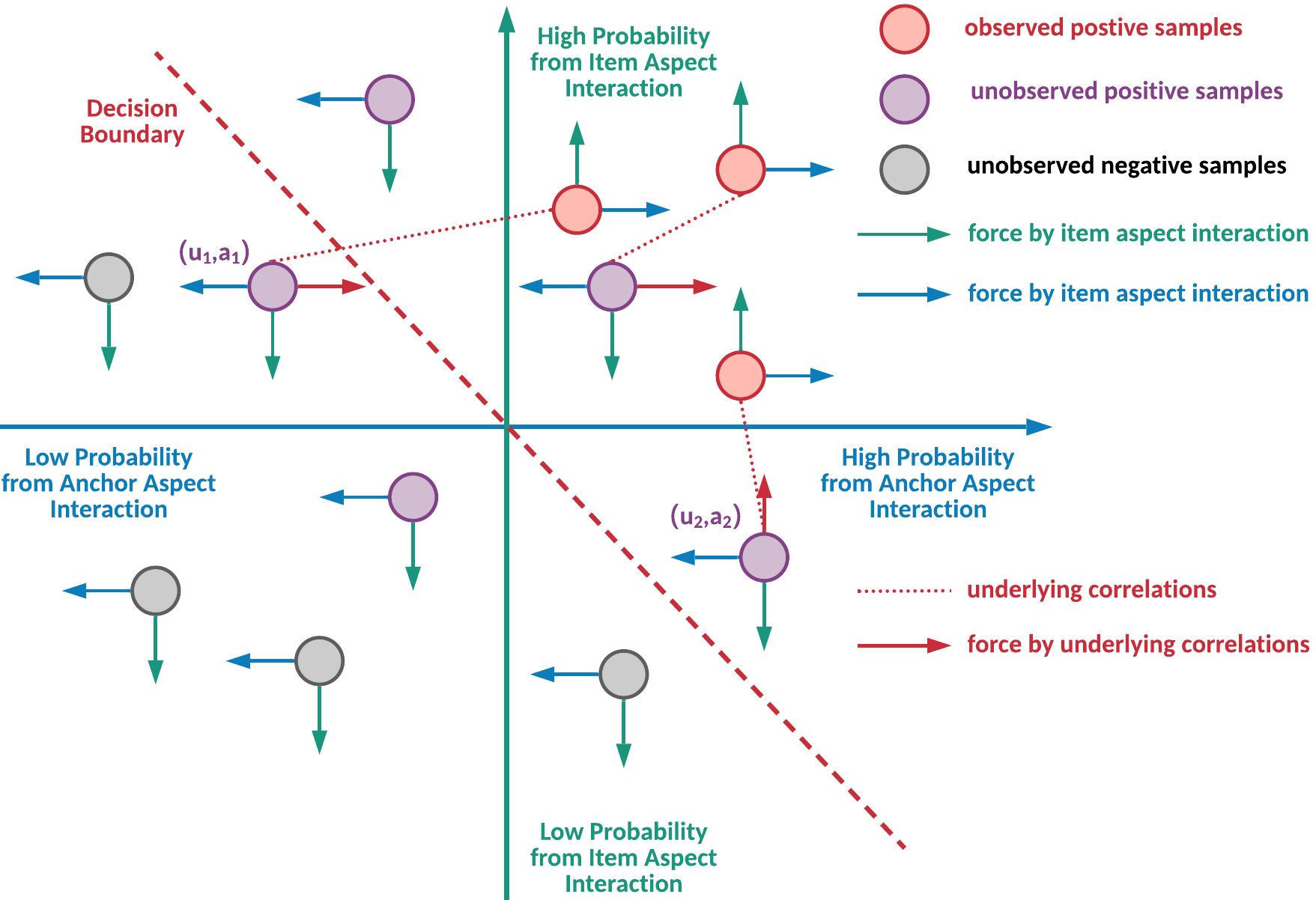}
	\vspace{-6mm}
	\caption{
		An illustrated example of training for TWINS with item aspect and anchor aspect interactions, where the original position of each sample represents the probability of receiving positive feedbacks from users (e.g., click) solely according to $\bm{y}^e_{pn}$, and arrows show the force from $\bm{y}^u_{pn}$ and $\bm{y}^a_{pn}$.
	}
	\label{fig:interaction}
	\vspace{-3mm}
\end{figure}

Fortunately, we reveal that TWINS is an effective solution to alleviate the issue above.
Compared with unobserved negative samples, we argue that unobserved positive samples are more likely to have correlations with observed positive samples.
Such correlations can either come from sharing similar anchors or similar items in users' browsing history $\mathcal{H}^u_p$ and anchor's broadcast history $\mathcal{H}^a_n$.
We argue that the former correlations can be captured by our anchor aspect interactions, and the latter ones can be modeled by our item aspect interactions.
Take Figure~\ref{fig:interaction} as an instance, where the original position of each sample represents the probability of receiving positive feedbacks from users solely governed by $\bm{y}^e_{pn}$, and arrows denote the force from $\bm{y}^u_{pn}$ and $\bm{y}^a_{pn}$.
We consider two specific user-anchor pairs $(u_1, a_1)$ and $(u_2, a_2)$.
Suppose that $a_1$ has a strong correlation with one of $u_1$'s desired anchors, then $\bm{y}^a_{pn}$ would push the sample to a relatively high probability from anchor aspect interaction.  
Similarly, assume that $u_2$'s browsed items is correlated with one of $a_2$'s broadcast items, then $\bm{y}^u_{pn}$ would push the sample to a relatively high probability from item aspect interaction.
Notably, these two samples are common among all unobserved samples.
Therefore, once the loss $\mathcal{L}$ that fuses all these information converges, the unobserved positive samples would be more likely to be located at the right or upper side of the decision boundary than the negative ones.

\section{Dataset Description}
\label{app:dataset}
We provide the detailed description for three real-world public benchmark datasets as follows.
\begin{itemize}[topsep = 3pt,leftmargin =5pt]
		\item \textbf{Yelp business dataset}\footnote{\url{https://www.yelp.com/dataset/documentation/main}} is formed of Yelp business data recording business interactions among businessmen.
		It consists of around 35,943,096 interactions among 1,233,453 businessmen through more than 160,310 business cases. 
		We treat the interaction between two businessmen as the interaction between a user and an anchor.
		And we regard those business cases as items.
		The average sequence length of browsing logs of users and anchors is 5.17.
		\item \textbf{Trust statement dataset}\footnote{\url{http://www.trustlet.org/downloaded_epinions.html}} is collected by Paolo Massa from the Epinions.com Web site in a 5-week period from November, 2003 to December, 2003. 
        It contains the logs of 49,290 people who rated a total of 139,738 different documents at least once, writing 664,824 reviews and 487,181 issued trust statements. People and documents are represented by anonimized numeric identifiers.
		We treat the trust record between two people as the interaction between a user and anchor.
		And we regard those documents as items.
		The average sequence length of browsing logs of users and anchors is 16.55.
		\item \textbf{Aminer citation dataset}\footnote{\url{https://www.aminer.cn/citation}} \citep{tang2008arnetminer} is extracted from DBLP, ACM, MAG (Microsoft Academic Graph), and other sources. 
		It contains 622,196 authors writing 243,266 papers, and 10,368,942 citations.
		Each paper is associated with abstract, authors, year, venue, and title.
		We treat the citation-relation between two authors as the interaction between a user and an anchor.
		And we regard those papers as items.
		The average sequence length of browsing logs of users and anchors is 2.33.
	\end{itemize}
We don't use some widely adapted e-commerce datasets created by Alibaba or Amazon, because they only can provide the sequential data in other sides.
More specifically, these datasets such as Tmall dataset\footnote{\url{https://tianchi.aliyun.com/dataset/dataDetail?dataId=42}}, Taobao E-Commerce dataset\footnote{\url{https://tianchi.aliyun.com/datalab/dataSet.html?dataId=408}}, Alipay dataset\footnote{\url{https://tianchi.aliyun.com/dataset/dataDetail?dataId=53}} only contain user's browsing logs (i.e., the sequential data in user side), which are definitely not suitable to simulate the live broadcast  triple-object interaction cases.
Notice that although some entities in the above dataset can be used as either users or anchors, our model will not reduce the one-side architecture, because the two-side architecture of TWINS is asymmetric, as we involve the interactive network from user aspect.

\begin{figure}[t]
	\centering
	\includegraphics[width=1.0\linewidth]{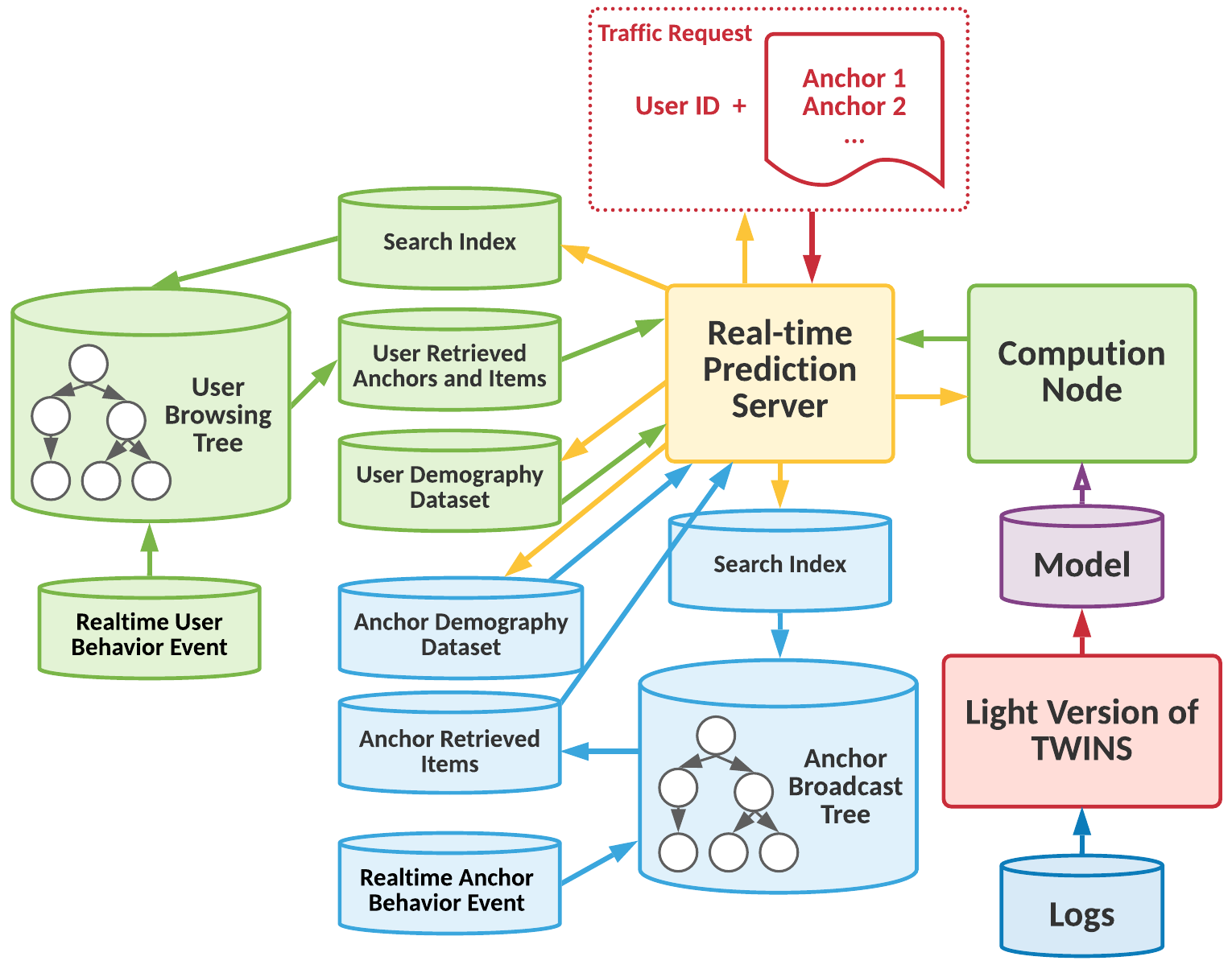}
	\vspace{-7mm}
	\caption{
		Online live broadcast recommender system with proposed TWINS model that partly shares similar idea with \citep{pi2020search}.
		The new system lightens the TWINS model, and builds tree structures for user browsing histories and anchor histories in a offline fashion to save computation and latency costs for online serving.
	}
	\label{fig:deploy}
	\vspace{-3mm}
\end{figure}

\section{Experimental Configuration}
\label{app:config}
We randomly split each dataset into training/validation/test sets at 6:2:2.
The learning rate is decreased from the initial value $1\times 10^{-2}$ to $1\times 10^{-6}$ during the training process.
The batch size is set as $2000$.
The weight for L2 regularization term is $4\times 10^{-4}$.
The dropout rate is set as $0.5$.
The dimension of embedding vectors is set as $64$.
The length of co-retrieval is set as $10$.
All the models are trained under the same hardware settings with 16-Core AMD Ryzen 9 5950X (2.194GHZ), 62.78GB RAM, NVIDIA GeForce RTX 3080 cards.

\section{Deployment Discussion}
\label{app:deploy}
In this section, we introduce our hands-on experience of deploying TWINS in the live broadcast recommender system in Alibaba.
As industrial recommender or ranker systems are required to response to massive traffic requests in a short time interval (e.g., one second \citep{pi2020search}), then the storage and latency constraints would become the main bottleneck for deploying existing search-based model \citep{pi2020search} and sequential model \citep{pi2019practice} to the online system.
We here develop two kinds of techniques to light the current TWINS model and develop an effective data structure, and introduce a new online live broadcast recommender system in Figure~\ref{fig:deploy}.
We further show the details as follows.

\minisection{Light Version of Module}
As the main computation costs come from the RNN model (as shown in Eq.~(\ref{eqn:rnn})) and the bi-attention model (as shown in Eqs.~(\ref{eqn:biattentioni}) and (\ref{eqn:biattentiona})), we tweak the original version of the TWINS model to obtain its light version.
Specifically, for the RNN model, inspired by LightRNN \citep{li2016lightrnn}, we use 2-Component (2C) shared embedding
for item representations; while for the bi-attention model, we remove $\bm{e}^u_p$ and $\bm{e}^a_n$ to reduce the computation costs.

\minisection{Tree Structure of Data}
Following the main idea of the implementation part in \citep{pi2020search}, we build two-level structured index for each user and anchor, which we call as user browsing tree and anchor broadcast tree respectively as illustrated in Figure~\ref{fig:deploy}.
More concretely, these trees follow the Key-Key-Value data structure where the first key is user id, the second keys are category ids of browsed items, and the last values are the specific behavior items that belong to each category.
For each user-anchor pair, we take the categories of the common ones in user's browsed items and anchor's broadcast items. 
After applying the proposed co-retrieval mechanism, the length of user's and anchor's item sequences can be significantly reduced which indeed releases much storage pressure in the online system.
Besides, these tree structures can be pre-built in an offline manner.

\end{document}